\documentclass[preprint,aps,nofootinbib]{revtex4-1}
\usepackage{bm}
\usepackage{epsf}
\usepackage{amssymb}
\usepackage{graphicx}
\usepackage{epstopdf}    
\usepackage{dcolumn}
\usepackage{amsfonts}
\usepackage{amsmath}
\usepackage{amssymb}
\usepackage{adjustbox}
\usepackage{bm}
\usepackage{color}
\usepackage{comment}
\usepackage{rotating}
\usepackage{adjustbox}
\newcommand{\edc}{\end{document}}
\newcommand{\bb} {\begin{thebibliography}}
\newcommand{\eb}{\end{thebibliography}}
\newcommand{\bi}[1]{\bibitem{#1}}
\newcommand{\bc}{\begin{center}}
\newcommand{\ec}{\end{center}}

\newcommand{\be}{\begin{equation}\small}
\newcommand{\ee}{\end{equation}\normalsize}
\newcommand{\bea}{\begin{eqnarray}}
\newcommand{\eea}{\end{eqnarray}}
\newcommand{\ba}{\begin{array}{l}   }
\newcommand{\lab}[1]{\label{#1}}
\newcommand{\ea}{\end{array}}

\newcommand{\dsfrac}{\displaystyle\frac}
\newcommand{\ds} {\displaystyle}
\newcommand{\summa}{\ds\sum}
\newcommand{\dssum}{\summa}
\newcommand{\re}[1]{(\ref{#1})}

\newcommand{\ci}{\cite}

\newcommand{\sigtil}{\tilde{\sigma}}

\newcommand{\dsint}{\ds\int}

\newcommand{\ti}{\tilde}
\def\bfr{{\bf r}}
\def\bfk{{\bf k}}

\newcommand{\vecr}{ \vec{r}}

\newcommand{{\vergul}}{  ,}

\newcommand{\veps}{\varepsilon }

\newcommand{\rhozero}{\ds{ \rho_0}}
\newcommand{\rhoone}{\ds{\rho_1}}

\newcommand{\Ek}{E_k}
\newcommand{\vecnab}{\mbox{\boldmath $\nabla$}}

\newcommand{\tpsi}{\tilde\psi}

\begin{document}
\draft
\title{ Tan's contact as an indicator of completeness and self-consistency of a theory.}
\author{Abdulla Rakhimov},\email{rakhimovabd@yandex.ru}
\address{
Institute of Nuclear Physics, Tashkent 100214, Uzbekistan
}


\begin{abstract}
It is well known that, Tan's contact could be calculated by using any of following three
methods: by the asymptotic behavior of momentum distribution;
 by Tan's adiabatic sweep theorem; or by the operator product expansion as an expectation value of the interaction term.
 We argue that, if a theory describing Bose (or Fermi) system with the only contact interaction
 is self consistent, then it should lead to the same result in all three cases. As an example we
  considered MFT based approaches and established that among existing approximations of MFT, the
 Hartree - Fock - Bogoliubov (HFB) approach is the most self consistent.  Actually, HFB  is able to describe existing experimental data on Tan's contact for dilute Bose gas, but fails to predict
 its expected behavior  at large gas parameter $(\gamma > 0.015)$. So, for appropriate description
 of properties of a Bose gas even at zero temperature, this approximation needs to be
 expanded by taking into account fluctuations in higher order then the second one.

\end{abstract}


\maketitle

\section{Introduction}
 The experimental discovery of superfluidity of ${}^{4}{He}$
  at very low temperatures and atmospheric pressure
 and realization of Bose- Einstein condensation (BEC) of alkali atoms \ci{anderson1995} have given impact to development of various theories describing thermodynamics of the system of ultra cold atoms. Most of them are based
 on field theoretical approaches, developed before for high energy physics, and nicely reviewed by Andersen \ci{Andersen}.
 In accordance with the classification  proposed by N. Proukakis and B. Jackson, the existing theoretical  formalisms may be classified, loosely speaking, into three "classes" of approaches, based on certain common conceptual notions shared between them \ci{tutorial}. Namely, mean field theories (MFT), number - conserving
 perturbative treatments and stochastic approaches. Although there is no universally accepted 'optimal' theory for description of ultra cold Bose gases at low temperatures, a researcher may  prefer one of those approaches depending on the nature of his main goal. For example, when the dynamics or the behavior of the system at critical point is not the issue, then mean field theory seems to be optimal.

 Mean field approaches for ultra cold gases rely on spontaneous symmetry breaking, which mathematically
 manifests itself by splitting of the Bose field operator $\psi(\bfr,t)$ into a mean field condensate contribution $\phi(\bfr,t)$ and an operator describing fluctuations (quantum, thermal) about this mean field.
 After such splitting, also called as Bogoliubov  shift \ci{Yukalovann} the full system hamiltonian breaks down
 into various contributions as $H=H_0+H_1+H_2+H_3+H_4$ based on the number of condensate and non- condensate factors contained in  each of them. For example, $H_0$ has no operators, while, $H_4$ includes fluctuation operators in fourth order.  Further, various approaches within MFT
 arise depending on the way of taking into account those fluctuations, since even simple $\lambda \phi^4$
 model has no analytical complete solution. For example, for weakly interacting ultra cold Bose gases, characterized by a small diluteness  gas parameter $\gamma$ , one limits himself to simple Bogoliubov or to the one loop
 approximation, when only $H_2$ term is taken into account. All and all , in general,  each approximation in the framework of MFT leads to a closed system of equations with respect to self energy, condensed fraction etc. Obviously,
 these equations should be solved self-consistently, which requires the self-consistency of a chosen  approach or a theory as a whole  by itself.

 In present work we propose that, evaluation of Tan's contact of a system with contact s-wave interaction
 may serve as a check point for the self consistency of a model. As an example we shall consider various approximations within MFT and check their self- consistency by evaluation of Tan's contact in different ways.

 Nearly fifteen years ago Shina Tan introduced \ci{tan1,tan2,tan3} a new
 quantity, $C$ which is further referred in the literature  as a Tan's contact. By using rigorous mathematical methods to study the system of fermions with contact interaction he obtained exact universal relations, which include the contact. He proved that, this quantity measures the density of pairs at short distances and determines the exact large momentum or high frequency behavior of various physical observables.
 Further, Tan's ideas were developed in works \ci{brprl100,brprl104,brprl106,lang2017,werner,combescot}
 and his relations have been rederived  and extended by using alternative methods. Particularly,
 Combescot et al. \ci{combescot} have shown that Tan's relations are valid not only for fermions, but also for bosons.
 It is remarkable that Tan's relations, including $C$, hold for any state of the system, few - body or many-body,
 homogenous or in a trapping potential, superfluid or normal, zero or nonzero temperatures \ci{brprl100}.

 Summarizing, Tan's contact for Bosons with zero range interaction may be theoretically evaluated (or measured experimentally) by using any of following equations \ci{pitbook}:
 \begin{itemize}
 \item
 By the asymptotic behavior of momentum distribution $n_k$
 \be
 C_n=\lim_{k\rightarrow \infty} k^4 n_k
 \lab{cn}
 \ee
 where $n_k$ is normalized to the total number of particles N, such that $\sum_{k}n_k=N$.
 So\footnote{Here and below we adopt $\hbar=1$  and $k_B=1$ for convenience}, $C$ has a dimensionality $length^{-4}$. .
\item
By Tan's adiabatic sweep theorem as
\be
C_E=\dsfrac{8\pi m a^2}{V}\left(\frac{\partial E}{\partial a}\right)
\lab{ce}
\ee
where $E$ and $V$ are the total energy and volume of the system, respectively, $a$ is the s- wave scattering length and $m$ is the mass of  particle. This equation manifests the relation between macroscopic thermodynamical
parameter $E$ and microscopic parameter $a$. Consequently, the variation of the total energy can be written
in following general form \ci{pitbook,ourjt}:
\be
dE=TdS-PdV+\mu dN+HdM+\frac{C V}{8\pi m a^2}da
\lab{DE}
\ee
In this sense Tan's contact and the scattering length may be considered as a conjugate parameters
of a  system regardless in the superfluid or normal phase.
\item
By the operator product expansion as an expectation value of the interaction term as \ci{brprl100}
\be
C_\psi=\dsfrac{(mg)^2}{V} \int d\bfr \langle \psi^{+}(\bfr)\psi^{+}(\bfr) \psi (\bfr) \psi(\bfr)\rangle
\lab{cpsi}
\ee
where $g=4\pi a/m$ is the coupling constant of zero range interaction.
 \end{itemize}
 From Eq.s \re{cn}-\re{cpsi} it is seen that for the case of quantum particles with point like interactions,
 short range correlations are embedded in Tan's contact, which is proportional to the probability that two
 particles approach each other very closely.

 Obviously, regardless of the way of evaluation (or measuring) of Tan's contact by using any of three
 equations \re{cn}, \re{ce} or \re{cpsi} one is supposed to obtain the same value i.e.
 \be
 C=C_n=C_E=C_\psi
 \lab{cnepsi}
 \ee
This trivial statement gives an opportunity to check self consistency of an applied theory. In the first part of
the present work we shall derive explicit expressions for $C$ and revise various versions of MFT in this way.
In the second part  of the work we shall compare our results with existing experimental data on $C$, and make an attempt to predict its behavior at large gas parameter.

Presently Tan's contact has been experimentally studied not only for fermions \ci{sagi,stewart}, but also for  bosons at ultra cold temperatures \ci{wild,makotyn,chang,fletcher}. Particularly, Tan's relations on tails of the momentum distribution and the tail of the transition rate have been tested experimentally by using short time probes of ultra cold atoms. Moreover, $C$ plays an important role in the radio frequency (rf) spectroscopy
\ci{brprl100}. As it is expected the values of Tan's contact, obtained from both kind
of measurements, ballistic and rf spectroscopy show a good agreement \ci{stewart}.

The present work is organized as follows. In Sect.2 we derive explicit expressions
for Tan's contact in various approaches of MFT, in Sect. 3  we shall study  self consistency of each approach
by numerical analysis and compare our theoretical predictions with experimental values of $C$. In the last section we present our conclusions. The details of calculations and summary of working equations are presented in Appendices A and B , respectively.

\section{Tan's contact for homogenous Bose gas in MFT}

A grand canonical ensemble of Bose particles with a short range s - wave interaction
is governed by the Euclidian action \ci{Andersen},
\begin{eqnarray}
S[\psi,\psi^{+}]=\dsint_{0}^{\beta}d\tau\dsint d\vecr
{\Large\{}
\psi^{+}(\tau,\vecr)[\partial_\tau-\dsfrac{\vecnab^2}{2m}-\mu]\psi(\tau,\vecr)
\nonumber \\
 + \dsfrac{g}{2}[\psi^{+}(\tau,\vecr)\psi(\tau,\vecr)]^2
{\Large\}},
\lab{stot}
\end{eqnarray}
where $\psi^{+}(\tau,\vecr)$ is a complex field operator  that creates  a boson
at the position $\vecr$, $\mu$  is the chemical potential,
  $\beta=1/T$  the inverse of temperature $T$.  This corresponds to the Hamiltonian
  \begin{eqnarray}
H&=&\int d\vec{r}\left\{\psi^+\left[-\dsfrac{\vecnab^2}{2m}-\mu\right]\psi+
\frac{g}{2}(\psi^+\psi)^2\right\}
\label{ham}
\end{eqnarray}
The quantities, required for evaluation of Tan's contact can be obtained by using following expressions:
 \be
 \ba
 \rhoone= \langle \tilde\psi^+ \tilde\psi\rangle=\dsfrac{1}{V}\summa_k n_k, \quad \quad F=\Omega+\mu N,
 \quad E=F+TS\\
 \Omega=-T\ln Z , \quad
Z=\ds{\dsint \cal{D} \psi \cal{D} \psi^{+} }\exp\{-S[\psi,\psi^{+}] \},\\
\langle (\tilde\psi^+ \tilde\psi)^2\rangle=
\rhozero^2+\rhozero\dsint d\bfr [3\langle \psi_{1}^2\rangle+\langle \psi_{2}^2\rangle]
+\frac{1}{4}\dsint d\bfr [\langle \psi_{1}^4\rangle+2\langle \psi_{1}^2 \psi_{2}^2\rangle+
\langle \psi_{2}^4\rangle]
 \lab{eqstoch}
 \ea
 \ee
 where $\rhoone$ is the density of uncondensed atoms, $\Omega$ - free  energy, $S$ is the entropy, $\rhozero$ is the condensed fraction introduced by standard Bogoliubov shift
 \be
 \psi(\tau,\bfr)=\sqrt{\rhozero}+\tpsi(\tau,\bfr)
 \lab{shift}
 \ee
and $\psi_1$, $\psi_2$ are the components of fluctuation field defined as
\be
\tpsi=\dsfrac{1}{\sqrt{2}}(\psi_1+i\psi_2), \quad \tpsi^+=\dsfrac{1}{\sqrt{2}}(\psi_1-i\psi_2)
\lab{psi12}
\ee
After the insertion of Eq.s \re{psi12}
into \re{stot} the total effective action is separated as follows
\be
\ba
S=S_0+S_1+S_2+S_3+S_4\\
S_0=\dsint_{0}^{\beta}d\tau\dsint d\bfr
{
\Large\{}
-\mu\rhozero+\dsfrac{g\rhozero^2}{2}
{\Large\}
},\\
S_1=\dsint_{0}^{\beta}d\tau\dsint d\bfr
{
\Large\{}
\sqrt{2\rhozero}(g\rhozero-\mu) \psi_1
{\Large\}
}
,\\
S_2=\dsfrac{1}{2}\dsint_{0}^{\beta}d\tau\dsint d\bfr
{
\Large\{}
[\partial _\tau-\dsfrac{\vecnab^2}{2m}-\mu+3g\rhozero] \psi_{1}^2+
[\partial _\tau-\dsfrac{\vecnab^2}{2m}-\mu+g\rhozero] \psi_{2}^2
{\Large\}
}
,\\
S_3=\dsfrac{g\sqrt{\rhozero}}{\sqrt 2}\dsint_{0}^{\beta}d\tau\dsint d\bfr
{
\psi_1[ \psi_{1}^2+\psi_{2}^2]
}
,\\
S_4=\dsfrac{g}{8}\dsint_{0}^{\beta}d\tau\dsint d\bfr
{
\Large\{}
\psi_{1}^{4}+2 \psi_{1}^2\psi_{2}^2+\psi_{2}^4
{\Large\}
}
\lab{s1234}
\ea
\ee
The Eqs. \re{stot}-\re{s1234} are exact equations of MFT for a homogenous Bose gas and can not be evaluated
exactly. The problem is hidden in the evaluation of the  path integrals
 over the fluctuating fields: It is well known that
"there is no handbook of path integrals", so one has to use an approximation. The only case when the path integral
can be evaluated explicitly is so called Gaussian integral, based on following formula \ci{faddeev}
\footnote{Here, there is a summation over repeated indices, $(a,b=1,2)$.}
\be
\ba
\int {\cal D}  \psi_1 {\cal D}  \psi_2 e^{-\frac{1}{2}\int dx dx'
	\psi_a(x)G_{ab}^{-1}(x,x')\psi_b(x')} e^{\int dx j_a(x)\psi_a(x)}\\
=(\sqrt{DetG}) \exp
{\left[\frac{1}{2}\int dx dx' j_a(x) \bar{G}_{ab}(x,x')j_b(x') \right]}
\lab{formula}
\ea
\ee
where $ x=(\tau,\bfr)$ , and $\bar{G}_{ab}(x,y)=[G_{ab}(x,y)+G_{ba}(y,x)]/2$ is usually  interpreted as
a Green function.

\subsection{Gaussian approximation}

As a first approach we limit ourselves to the case when in Eq.s \re{s1234} the terms $S_3$ and $S_4$ are neglected
 \footnote{In quantum field theory this corresponds to the one loop approximation.}.
From explicit expression for $S_2$ in \re{s1234} one obtains the propagator $G_{ab}(x,x')=(1/V\beta)
\summa_{n,k}G_{ab}(\omega_n,\bfk)\exp(i\omega_n(\tau-\tau')+i\bfk (\bfr-\bfr'))$, where in momentum space
\begin{eqnarray}
G(\omega_n,\bfk)=\dsfrac{1}{\omega_{n}^{2}+\Ek^2}\left(
\begin{array}{cc}
\varepsilon_k+g\rhozero-\mu& \omega_n\\
\nonumber
 -\omega_n&\varepsilon_k+3g\rhozero-\mu
\end{array}\right),\\
 \\ \nonumber
\lab{propagator}
\end{eqnarray}
  $(a,b=1,2) $ and  $\omega_n=2\pi nT$ is the Matsubara frequency, $\Ek=\sqrt{\varepsilon_k+3g\rhozero-\mu}\sqrt{\varepsilon_k+g\rhozero-\mu}$ is   the quasiparticle  (Bogolon) dispersion
 with
$\varepsilon_k=\bfk^2/2m$. Now using Eqs. \re{eqstoch} , \re{formula} and \re{Aident}
 leads to the following
free energy at zero temperature
\be
\Omega(T=0)=-V\mu\rhozero+\frac{Vg\rhozero^2}{2}+\frac{1}{2}\summa_k (E_k-\veps_k)
\lab{omega}
\ee
 In a stable equilibrium, this should be minimized with respect to $\rhozero$ to give:
 \be
 \ba
 \dsfrac{\partial \Omega}{\partial \rhozero}=-V\mu+Vg\rhozero=0,
 \\
 \mu=g\rhozero
 \lab{minomega}
 \ea
 \ee
 Now inserting this chemical potential into $E_k$  one obtains
 a linear  at low momentum  dispersion
 \be
 E_k=\sqrt{\veps_k}\sqrt{\veps_k+2g\rhozero}=ck+O(k^3),
 \lab{dispgs}
 \ee
  with
 the sound velocity  $c=\sqrt{g\rhozero/m}$ .

 For the condensate depletion $\rhoone$ at zero temperature it is easy to obtain following equation
 \be
 \rhoone(T=0)=\frac{1}{2V}\summa_k \left[\frac{\veps_k+g\rhozero}{E_k}-1\right]\equiv \frac{1}{V}\summa_k n_k
 \lab{rho1}
 \ee
 where we used Eq.s \re{eqstoch}, and \re{Aident}, and hence
 \be
 C_n (Gaussian)= \lim_{k\rightarrow \infty} k^4 n_k=(gm\rhozero)^2=(4\pi a \rho)^2 n_0^2=C_{class} n_{0}^{2}
 \lab{cgauss}
 \ee
 where $n_0=\rhozero/\rho$ is the condensate fraction,  $\gamma=a^3 \rho$  is the gas parameter, and  $C_{class}=16\pi^2\gamma^2/a^4$ is the Tan's contact, corresponding to the case when all fluctuations have been neglected .
When the total number of particles (not the chemical potential) is fixed and given by the density $\rho$, the density of condensate
 $\rhozero$ in above equations can be found as a solution to the following equation
\be
\rhozero=\rho-\rhoone=\rho-\frac{1}{2V}\summa_k \left[\frac{\veps_k+g\rhozero}{\sqrt{\veps_k}\sqrt{\veps_k+2g\rhozero}}-1\right]=\rho-\frac{(mg\rhozero)^{3/2}}{3\pi^2}
\lab{eqrhozero}
\ee

Now we pass to calculation of $C_E$, defined by \re{ce}. First, using Eq.s \re{omega} and \re{minomega} we represent the total energy at zero temperature  as
\be
E=\Omega+\mu N=\frac{Vg\rho^2}{2}-\frac{Vg\rho_{1}^{2}}{2}+\frac{1}{2}\summa_k(E_k-\veps_k)
\lab{Etot}
\ee
Following the ideology of the  Gaussian approach, when the fluctuations, explicitly higher than  the first  order, are neglected we can rewrite the last equation as
\be
\ba
E=E_0+E_{fluc}, \quad \quad E_0=\dsfrac{Vg\rho^2}{2},\\
E_{fluc}=\dsfrac{V}{4\pi^2}\dsint_{0}^{\infty} k^2 dk (\sqrt{\veps_k}\sqrt{\veps_k+2g\rhozero}-\veps_k)
\lab{etotint}
\ea
\ee
The integral in Eq. \re{etotint} is divergent. This may be evaluated by using dimensional regularization \ci{ouryee} or just by subtracting infinite parts from the integrand, leading to the same result. So, using the method of subtraction one may easily obtain
\be
\ba
E_{fluc}=\frac{1}{2}\summa_k(E_k-\veps_k)\rightarrow \frac{1}{2}\summa_k[E_k-\veps_k-g\rhozero+
\frac{(g\rhozero)^2}{2\veps_k}]=
\frac{8Vm^{3/2}(g\rhozero)^{5/2}    }{15\pi^2}
\lab{efluc}
\ea
\ee
Taking the derivative with respect to $a$ requires an explicit expression for $d\rhozero/da$, which could be
obtained by differentiation of both sides of \re{eqrhozero} and solving it with respect to $d\rhozero/da$. This gives
\be
\frac{d\rhozero}{da}=-\frac{\rhozero}{a}\frac{1}{(1+\frac{\sqrt{\pi} } {4\sqrt{n_0\gamma}  })}
\lab{drho0}
\ee
Finally, by using Eq.s \re{ce}, \re{efluc} and \re{drho0} we obtain
\be
C_E(Gauss)= C_{class} \left[1+\dsfrac{64n_{0}^{5/2}\sqrt{\gamma}     }{3(\sqrt{\pi}+4\sqrt{\gamma n_0})     }\right]
\lab{cegauss}
\ee

As to the $C_\psi$, defined by \re{cpsi} it can be easily found from equations \re{eqstoch}
and \re{Aident} as
\be
C_{\psi}=C_{class} \left[1+2(n_1+\tilde{\sigma}) \right]
\lab{cpsigaus}
\ee
where $n_1=\rho_1/\rho$ . In Eq. \re{cpsigaus}  we neglected high order fluctuations and introduced the fraction of
anomalous density as   $\tilde{\sigma}=(\langle \tilde\psi^+\tilde\psi^++\tilde\psi\tilde\psi \rangle )/2\rho$.

  \subsection{Optimized Gaussian approximation}

  In the previous subsection we have taken into account the depletion $\rhoone$ and anomalous density $\sigma$
  only up to the linear order, neglecting the terms $S_3$ and $S_4$ in Eqs. \re{s1234}. Below we
   extend those relations for Tan's contact by taking into account quantum fluctuations in a more accurate way.
  For this purpose, we employ variational perturbation theory, developed by Stevenson long years ago
  \ci{stevenson} for the $\lambda\phi^4$ theory and further referred as a $\delta$ - expansion method \ci{pinto}. In this method one introduces an auxiliary parameter $\delta$ and uses a perturbative scheme in power   series of $\delta$, which is set to
  unity at the end of calculations. Note that, the main drawback of this theory
  is that, there is  an arbitrariness in the
choice of the expansion parameter $\delta$.

  For the effective action \re{stot} the method includes two variational parameters, which may be fixed
  by principle of minimal sensitivity.
    In present section we apply variational perturbation theory to derive explicit expressions for Tan's contact
  , limiting ourselves to the first order in $\delta$, which is referred in the literature as an  Optimized Gaussian Approximation. This will give us an opportunity to take into account
  $\rhoone$ as well as $\sigma$ up to the second order explicitly. Below we present the main equations, needed for calculation of Tan's contact, referring a reader to the Appendix A for  details. Note that, present approximation  is equivalent to Hartree - Fock - Bogoliubov (HFB) approach \ci{tutorial,Yukalovann} used in Hamiltonian formalism. The preference of
the path integral formalism is that, in contrast to Hamiltonians one, it gives a natural opportunity
  for going beyond HFB approximation, as it was shown by Stancu and Stevenson \ci{stancu}.

  Thus, for the free energy and densities we have \footnote{See Appendix A for the details}
  \be
  \ba
\Omega(T=0)=-N\mu+\dsfrac{Vg\rho^2}{2}
+\dsfrac{Vg(\rhoone^2-2\rhoone\sigma-\sigma^2)}{2}+
\frac{1}{2}\summa_k (E_k-\veps_k),\\
\rhoone(T=0)=\frac{1}{2V}\summa_k \left[\frac{\veps_k+\Delta}{E_k}-1\right]\equiv \frac{1}{V}\summa_k n_k=
\frac{(\Delta m)^{3/2}}{3\pi^2},\\
\sigma(T=0)\equiv
{\rho\tilde\sigma}=
-\frac{\Delta}{2V}\summa_k \left[\frac{1}{E_k}-\frac{1}{\veps_k}\right]=
\frac{(\Delta m)^{3/2}}{\pi^2}\approx \frac{\Delta m^{3/2}}{\pi^2}\sqrt{g\rhozero}
\lab{omegahfb}
\ea
\ee
where the energy dispersion is similar to the Bogoliubov one:
\be
E_k=\sqrt{\veps_k}\sqrt{\veps_k+2\Delta}
\lab{Ekhfb}
\ee
For the zero temperature energy, from the  relation $E=\Omega+V\rho\mu$, one obtains
\be
\ba
E(T=0)=\dsfrac{Vg\rho^2}{2}+
\dsfrac{V g}{2}[\rhoone^2-\sigma^2-2\rhoone\sigma]+
\dsfrac{1}{2}\summa_k (E_k-\veps_k-\Delta+\dsfrac{\Delta^2}{2\veps_k})=\\
\dsfrac{Vg\rho^2}{2}+
\dsfrac{Vg}{2}[\rhoone^2-\sigma^2-2\rhoone\sigma]+\dsfrac{8V\Delta^{5/2}m^{3/2}       }{15\pi^2}
\lab{etothfb}
\ea
\ee
where the subtraction terms were introduced. The Eq.s \re{omegahfb}-\re{etothfb} include a
key parameter $\Delta$, which may be found by the physical solution ($\Delta\geq 0 $) of following equation of MFT:
\be
\Delta=g(\rhozero+\sigma)=g(\rho-\rhoone+\sigma)
\lab{delta}
\ee
This equation gives following explicit expression for the derivative of $\Delta$ with respect to $a$ as
\be
\Delta'_{a}=\dsfrac{\Delta}{a}\dsfrac{1}{[1+6\pi a (\rhoone-\sigma)/m\Delta]}
\lab{dda}
\ee
which is needed for evaluation of $dE/da$ by using Eq.s \re{ce} and \re{etothfb}.
Therefore, in HFB approach we obtain following expressions for Tan's contact
\be
C_n=(\Delta m)^2=(c m)^4
\lab{cnnhfb}
\ee
\be
\ba
C_E=C_{class}(1+W_E),\\
W_E=n_\sigma+\Delta'_{a}\left[\dsfrac{2mn_1}{\pi\rho}+\dsfrac{3an_\sigma}{\Delta}\right]
\lab{cehfb}
\ea
\ee
\be
\ba
C_\psi=C_{class}(1+W_\psi),\\
W_\psi=2(n_1+\tilde\sigma-2n_1 \tilde\sigma)-n_\sigma
\lab{cepsi}
\ea
\ee
with $n_\sigma=n_{1}^2-\tilde\sigma^2-2n_1\tilde\sigma$.  From equation \re{cnnhfb} it is seen that,  Tan's contact $C_n$, calculated from the tail of density distribution, is  related to the sound velocity $c=\sqrt{\Delta/m}$ and may be  directly observed experimentally by
sound velocity measurements.

\subsection{Bogoliubov approach}

In Bogoliubov approximation \ci{bog} the energy dispersion and the total energy are given as
\be
\ba
E_k=\sqrt{\veps_k}\sqrt{\veps_k+2g\rho}\\
E(T=0)=\dsfrac{Vg\rho^2}{2}+
\dsfrac{1}{2}\summa_k (E_k-\veps_k-g\rho+\dsfrac{(g\rho)^2}{2\veps_k})=\\
\dsfrac{2V\pi \gamma^2}{ma^5}\left[1+\dsfrac{128\sqrt{\gamma}}{15\sqrt{\pi}}\right]
\lab{bogeq}
\ea
\ee
and hence
\be
C_E=C_{class}\left[1+\dsfrac{64\sqrt{\gamma}}{3\sqrt{\pi}}\right]
\lab{cebog}
\ee
Remarkably, the expression for the total energy in \re{bogeq} coincides with that one obtained
long years ago by Lee Huang and Yang (LHY)  \ci{leeyang} in hard core boson model, and the equation \re{cebog}
  for $C_E$ does with the result by Schakel \ci{schakel} derived in a similar way.

  The question arises, what is the difference between Gaussian and Bogoliubov approximations?
  The main difference is that in Gaussian approximation one has preliminary
  solve the equation \re{eqrhozero} with respect to $\rhozero$ for a given $\gamma$,
  while in Bogoliubov one there is no need to solve any equation. This fact makes  Bogoliubov approximation
  attractive and the most practical one in order to make a fast estimation of a physical quantity  in the BEC regime.

  Formally, Eq.s \re{bogeq} may be derived from HFB approach by setting there $\Delta=g\rho$,
  $\rhoone^2\rightarrow 0$ and $\sigma\rightarrow 0$ explicitly in Eq.s \re{omegahfb}-\re{etothfb}.
  So, particularly, one obtains

  \be
  \ba
  \rhoone(T=0)=\frac{1}{2V}\summa_k \left[\frac{\veps_k+g\rho}{\sqrt{\veps_k}\sqrt{\veps_k+2g\rho}}
  -1\right]\equiv \frac{1}{V}\summa_k n_k=
\frac{(g\rho m)^{3/2}}{3\pi^2},\\
    n_0=1-\dsfrac{\rhoone}{\rho}=1-\dsfrac{8\sqrt{\gamma}}{3\sqrt{\pi}}\\
     C_n=C_\psi=C_{class}
    \lab{rho1bog}
  \ea
  \ee
  From Eq. \re{rho1bog} one may conclude that, Bogoliubov approximation takes into account
  the gas parameter up to the first order in the expansion by $\sqrt{\gamma}$ in evaluation of the condensed fraction.
The net results of the  present section are summarized in Table 1 of Appendix B.

\section{Results and discussions}

Now we are in the position of studying three versions of MFT for self consistency in the spirit of
the requirement in Eq. \re{cnepsi}. In Figs.1 we present Tan's contact obtained in Bogoliubov (Fig.1a), Gaussian
(Fig.1b) and HFB (Fig.1c) approximations. Here dashed, solid and dotted curves correspond to $C_n$, $C_E$ and
$C_\psi$ defined by equations \re{cn}, \re{ce} and \re{cpsi} respectively.
From Fig.1a it is seen that Bogoliubov approximation satisfies the first  equality $C_n=C_\psi$, but does
not the second one, i.e. $C_\psi\neq C_E$. As to the Gaussian (one loop ) approximation
the difference between these three quantities is rather notable (see Fig.1b). In this sense, Bogoliubov approximation, seems more reliable than the Gaussian one. This fact can  explain popularity of Bogoliubov approximation, including LHY  terms \ci{leeyang} in the literature \ci{wild,petrov}.
From Fig.1c it is seen that the discrepancy between $C_n$, $C_E$ and
$C_\psi$ is  rather small for the Variational Gaussian approximation. Hence, one may  conclude
that HFB approximation can be regarded as the most complete and self consistent one among other existing
MFT based  approaches. Nevertheless, strongly speaking, HFB is also needed corrections, especially
for $\gamma>0.002$, arising from the high order quantum fluctuations. The intensity of such fluctuations is
almost proportional to the fraction of uncondensed particles $n_1$. As it is seen from Fig.1d
even at $\gamma\sim 0.005$ the depletion is about $15 \%$. Note that , in superfluid helium ${}^4$He,
$n_1\approx 90\%$.

On the other hand, one may judge about an appropriateness of any theory just by comparing
its predictions with experimental measurements. In Fig.2a we compare our predictions for Tan's contact
given by HFB approach with the experimental data on ${}^{85}$Rb atomic condensate at fixed density
$\rho=5.8\mu m ^{-3}$. It is seen that HFB approximation is able to describe $C$ rather satisfactory
up to $a/a_B< 1200 $, which corresponds to $\gamma\approx 0.0015$. Moreover, HFB predictions
for the Tan's contact is in a good agreement with path-integral ground - state (PIGS) Monte Carlo calculations
performed by Rossi and Salasnich  \ci{rossisalas}.

Unfortunately, presently Tan's contact for a Bose gas has been measured at very small
values of the gas parameter, $\gamma\leq 0.002$. To predict its behavior at larger $\gamma$
we calculated Tan's contact in the region $0\leq \gamma \leq 0.2$
\footnote{For superfluid helium  $\gamma\approx 0.6$  }
and presented the results in Fig.2b. It is seen that PIGS Monte Carlo method predicts
a smooth increasing of $C$, while the latter  remains practically
unchanged in HFB approximation for $\gamma > 0.05$

 \begin{figure}[h]
\begin{minipage}[h]{0.49\linewidth}
\center{\includegraphics[width=1.1\linewidth]{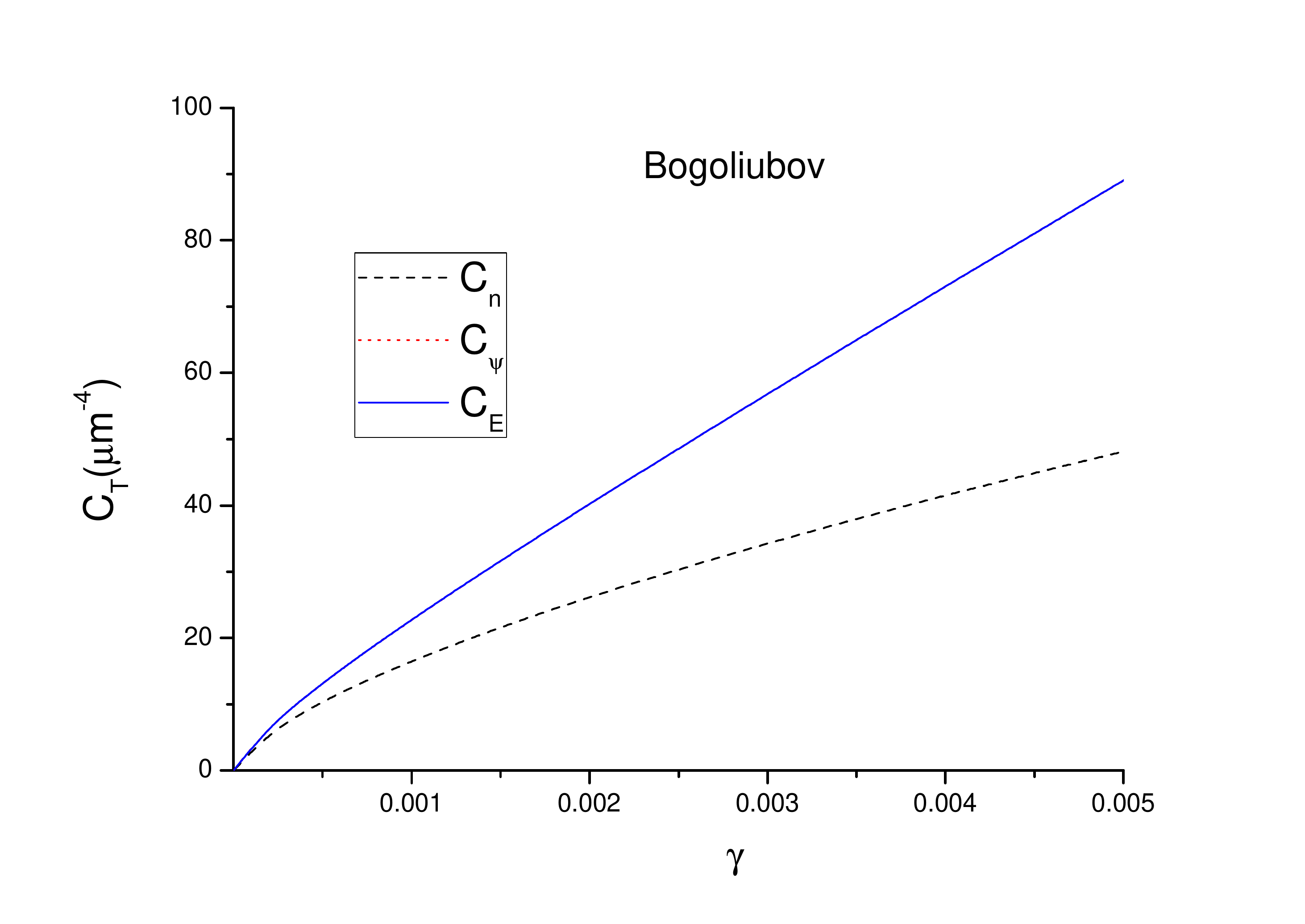}\\ a) }
\\
\end{minipage}
\hfill
\begin{minipage}[h]{0.49\linewidth}
\center{\includegraphics[width=1.1\linewidth]{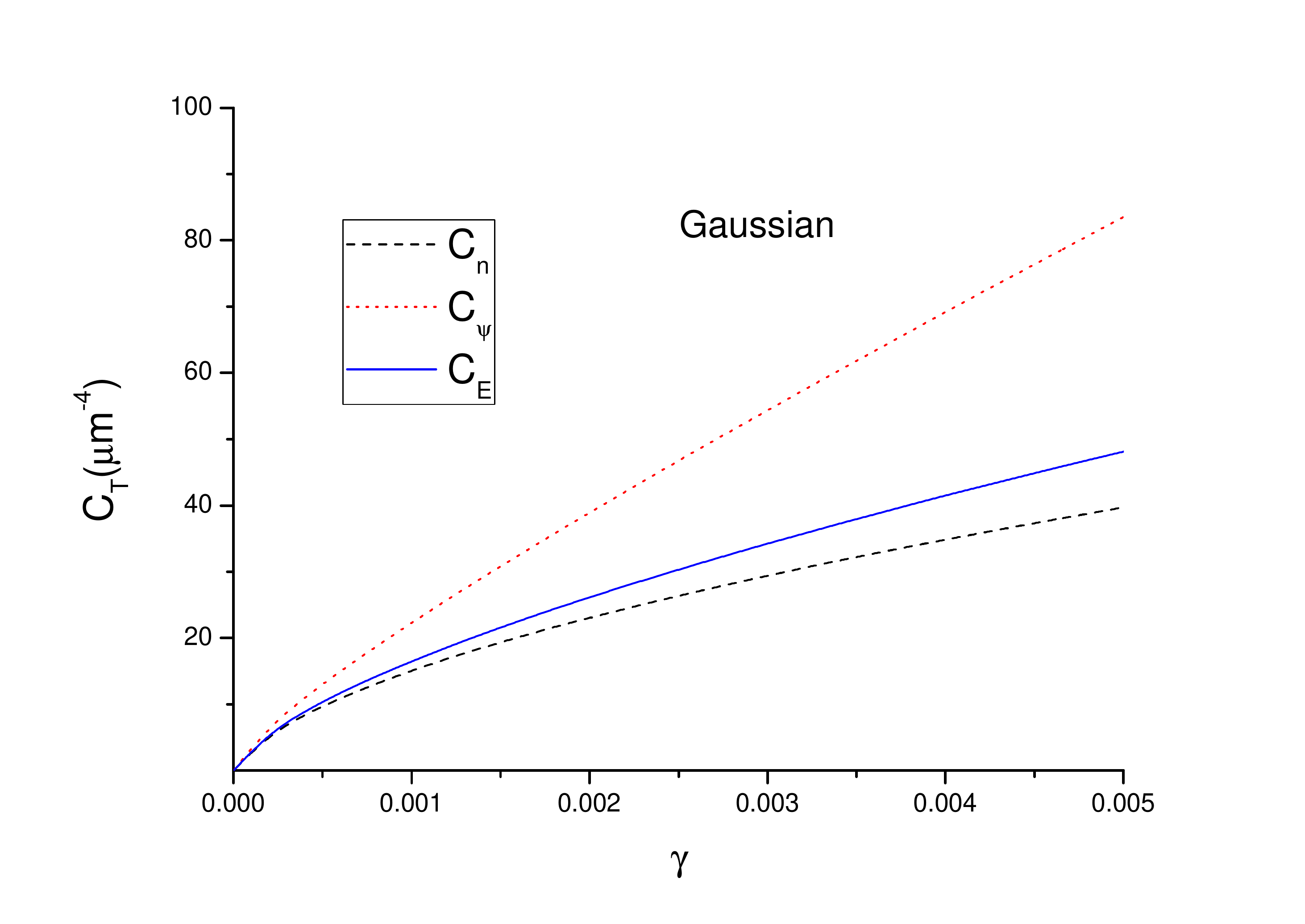} \\ b) }
\\
\end{minipage}
\medskip
\medskip
\medskip
\begin{minipage}[H]{0.49\linewidth}
\center{\includegraphics[width=1.1\linewidth]{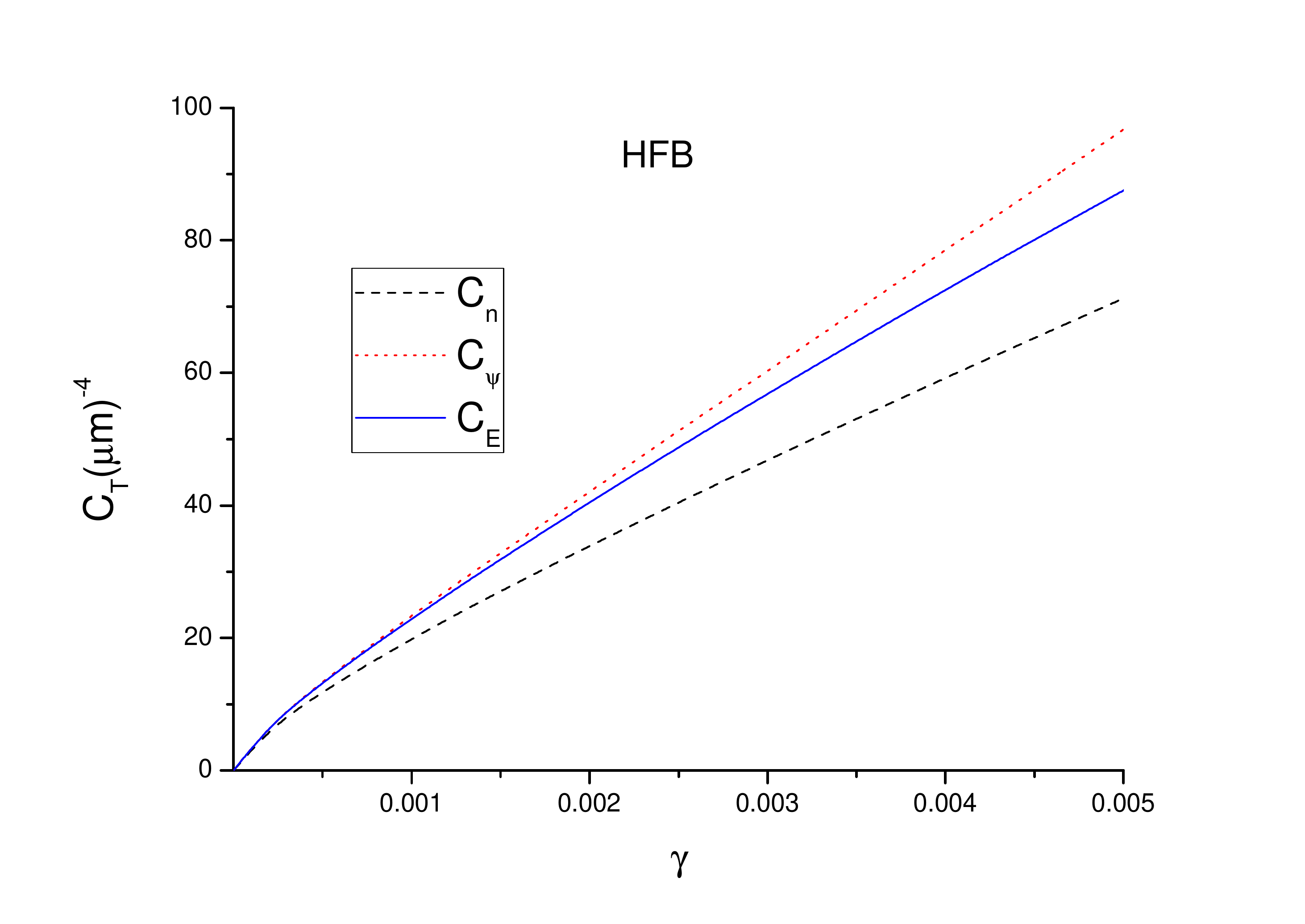}    \\ c)}
\\
\end{minipage}
\begin{minipage}[h]{0.49\linewidth}
\center{\includegraphics[width=1.1\linewidth]{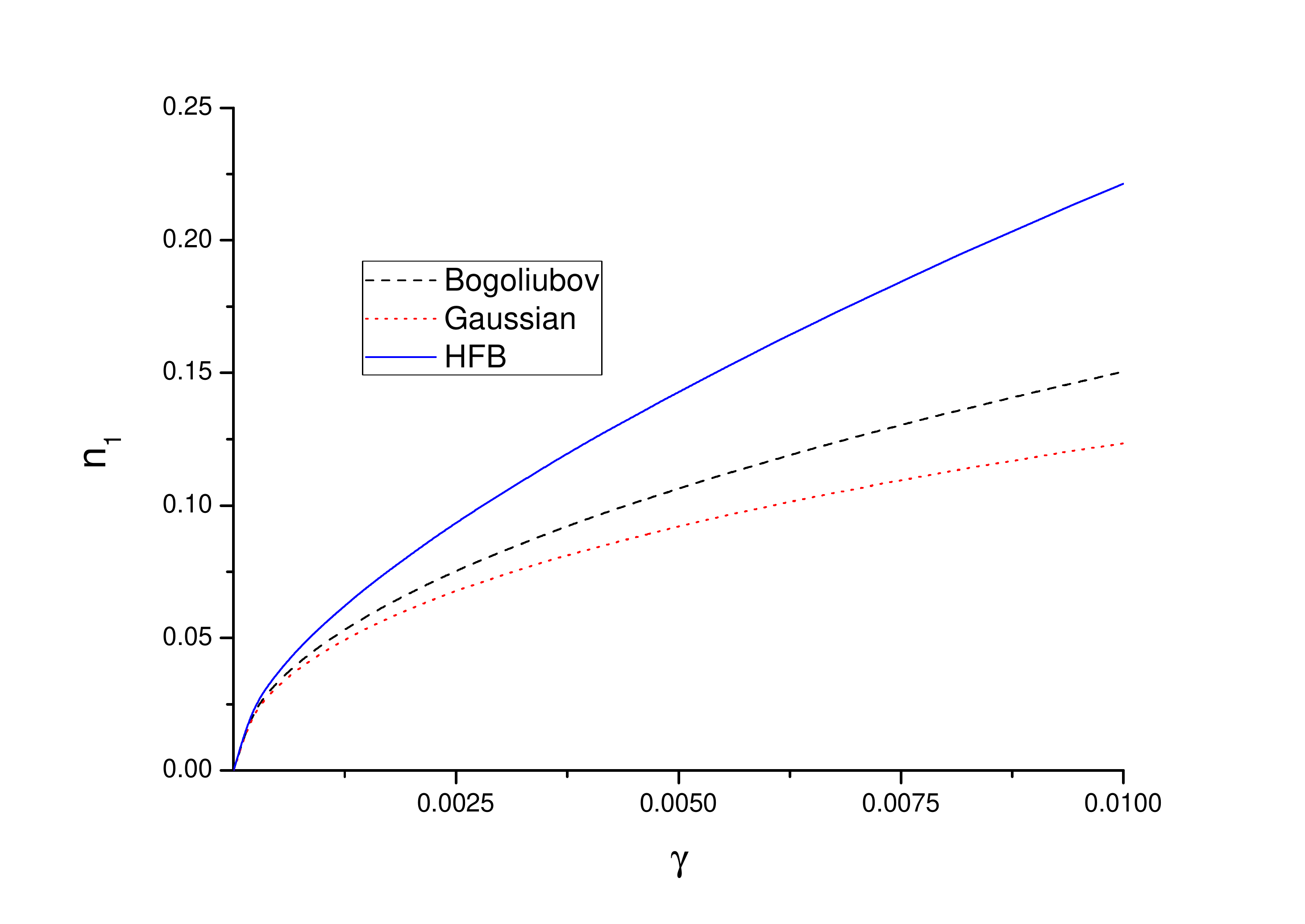} \\ d) }
\end{minipage}
\hfill
\caption
{
 Tan's contact  as a function of the gas parameter $\gamma=\rho a^3$ in Bogoliubov (a), Gaussian (b) and HFB approximations (c).
Dashed, solid  and dotted lines  are obtained with equations \re{cn} , \re{ce} and \re{cpsi},
 respectively. The corresponding condensate depletions, $n_1=N_1/N$,  are presented in Fig. 1 (d).
}
  \label{Fig1}
\end{figure}

 \begin{figure}[h]
\begin{minipage}[h]{0.49\linewidth}
\center{\includegraphics[width=1.1\linewidth]{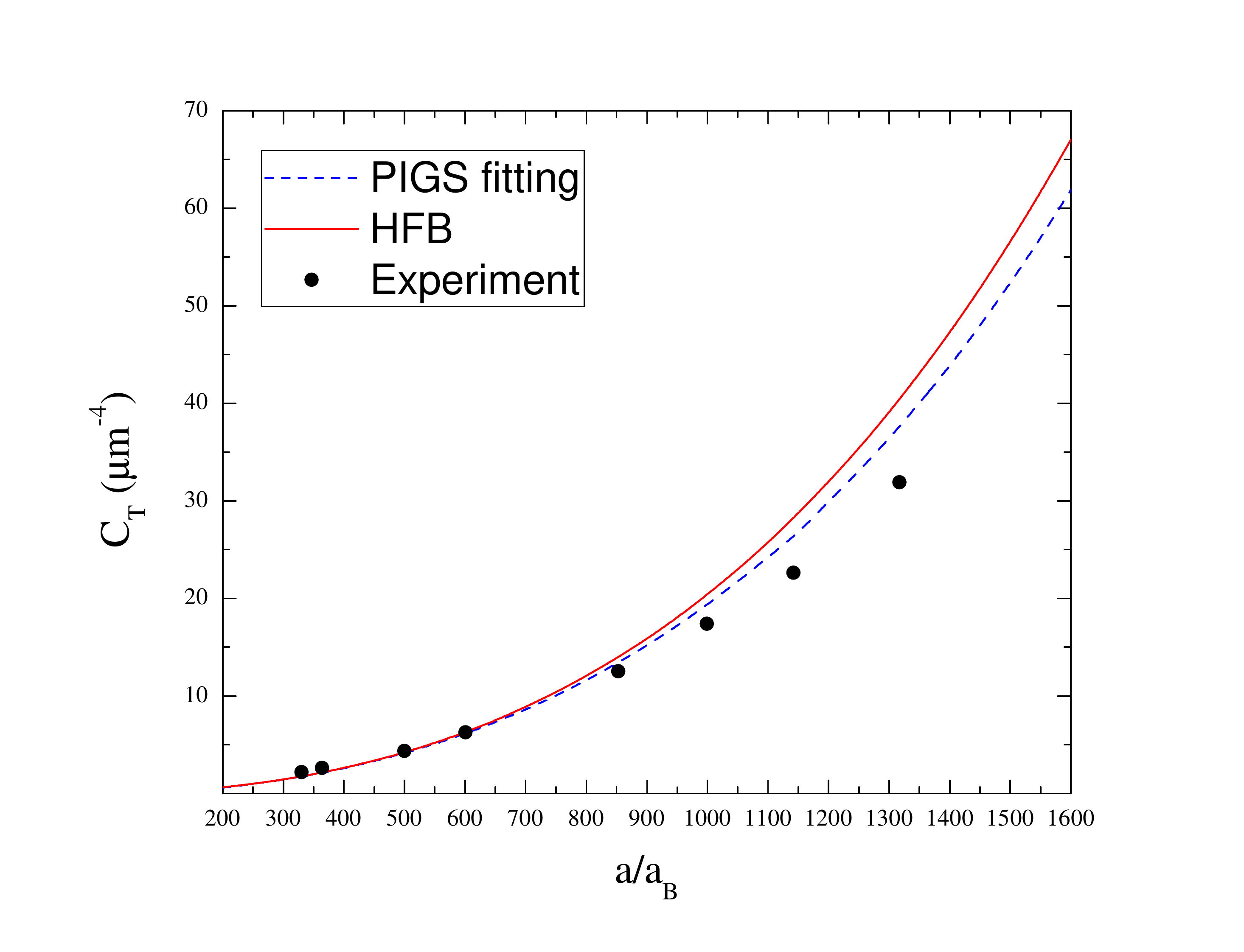}\\ a) }
\\
\end{minipage}
\hfill
\begin{minipage}[h]{0.49\linewidth}
\center{\includegraphics[width=1.1\linewidth]{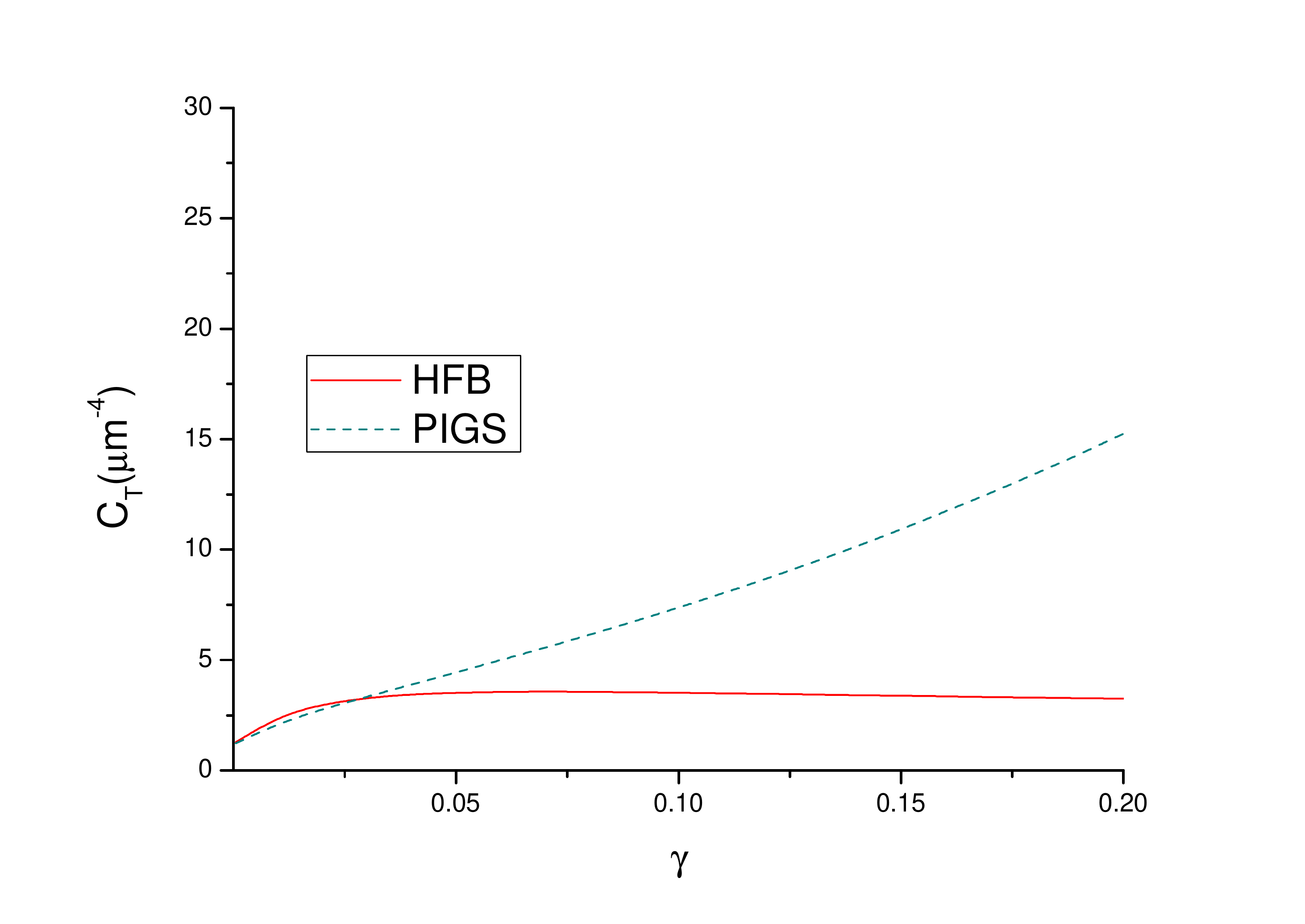} \\ b) }
\\
\end{minipage}
\hfill
\caption
{ (a):  The contact in natural units $(\mu m)^{-4}$ vs scattering length $a$ computed in
PIGS Monte Carlo method \ci{rossisalas} (dashed curve) and HFB approaximation (solid curve).
Filled circles are experimental data of  Wild et al. \ci{wild} obtained for
 ${}^{85}{Rb}$ atomic condensate. The scattering length $a$
is given in units of the Bohr radius $a_B=5.3 10^{-5} \mu m$; \\ (b): Tan's contact at large values of the gas parameter $\rho a^3$ in
HFB (solid line) and PIGS Monte Carlo (dashed line) \ci{rossisalas}. In both figures the density is fixed
in its typical value
as $\rho=5.8 \mu m ^{-3}$.
 }
  \label{Fig2}
\end{figure}

 \section{ Conclusion}

    We have  derived explicit expressions for Tan's contact of Bosons at zero temperature within various
    approximations based on mean field theory. Numerical analysis, made with these equations
    gave us an opportunity to study such approximations for completeness and self consistency. We have shown that
    in this concept Hartree - Fock - Bogoliubov approximation, derived within optimized Gaussian
    perturbation theory satisfies the requirement $C_n=C_E=C_\psi$ better than one loop or Bogoliubov approximations.

    Moreover, HFB  predictions are in a good agreement with existing experimental data as well as with
    Monte Carlo calculations for small values of the gas parameter. However, for large values of $\gamma$
    HFB needs serious corrections . These could be performed by extension of present approach
    in the spirit of Post Gaussian Perturbative  approximation, which includes the second order
    $\delta $ - expansion \ci{stancu}. It is expected that , such extension
    would give rise to a desired logarithmic term, which is used in the literature \ci{rossisalas,schakel,brnieto}.
    The work is on progress.

\section*{Acknowledgments}

    We are obliged to B. Tanatar for useful discussions.   This work is partly supported by Scientific and Technological Research Council of Turkey (TUBITAK) and Ministry of Innovative Development of the Republic of Uzbekistan.

\clearpage
\appendix
\section{Derivation of $\Omega$}
\label{sec:A}
In this appendix we present the derivation of the free energy $\Omega$ given in Eq.\, \re{omegahfb}. Inserting Eq.\, \re{shift} into the action Eq.\, \re{stot}, the latter can be divided into the following parts

\be
\ba

	S  =S_0 + S_1+S_2+S_3+S_4\\
	S_0 = \int_{0}^{\beta}d\tau\int d\vec{r}\left\{ -\mu \rho_0 + \dsfrac{g\rho_0^2}{2} \right\} \\
	S_1 = \int_{0}^{\beta}d\tau\int d\vec{r} \left\lbrace
\left[  g\rho_0^{3/2} -\mu \sqrt{\rho_0}\right] \ti{\psi} + h.c\right\rbrace \\
	S_2 = \int_{0}^{\beta}d\tau\int d\vec{r} \left\lbrace \ti{\psi}^+\left[ \partial_\tau - \frac{\nabla^2}{2m}\ +
2g\rho_0 -\mu\right]\ti{\psi} + \dsfrac{g\rho_0}{2}(\ti{\psi}^{+2} + \ti{\psi}^2) \right\rbrace\\
	S_3 =g \sqrt{\rho_0} \int_{0}^{\beta}d\tau \int d\vec{r}
\left\lbrace  \ti{\psi}^{+} \ti{\psi}^2 + \ti{\psi}^{+2} \ti{\psi} \right\rbrace\\
	S_4 = \dsfrac{g}{2}\int_{0}^{\beta} d\tau \int d\vec{r} \ti{\psi}^+ \ti{\psi}^+ \ti{\psi}\ti{\psi}.
	\label{eqS1234}
\ea
\ee

Now employing the $\delta$-expansion method, we add to the total action Eq. \re{eqS1234}, the term $ (1-\delta)\int_{0}^{\beta} d\tau \int d\vec{r} \left[ \Sigma_{n}(\ti{\psi}^+ \ti{\psi}) + (1/2) \Sigma_{an} (\ti{\psi}^{+}\ti{\psi}^{+} +\ti{\psi}\ti{\psi})\right] $ and make replacement   $g\rightarrow \delta g$. Then, after presenting $\ti{\psi} $ and $\ti{\psi}^+$ in Cartesian form as
\be
	\ba
	\ti{\psi} = \frac{1}{\sqrt{2}}(\psi_1 + i \psi_2).\\
	\ti{\psi}^+ = \frac{1}{\sqrt{2}}(\psi_1 - i \psi_2).
	\ea
\ee
the total action may be rewritten as follows \cite{ouraniz2part1}
\be
\ba
		S  =S_0 + S_{free}+S_{int}\\
	S_{free} = \frac{1}{2}\int_{0}^{\beta}d\tau\int d\vec{r}\left\lbrace i \epsilon_{ab} \psi_a \partial_\tau \psi_b +\psi_1 (-\frac{\nabla^2}{2m}+ X_1)\psi_1 + \psi_2(-\frac{\nabla^2}{2m}+ X_2)\psi_2 \right\rbrace.  \\
	S_{int} = S_{int}^{(1)} + S_{int}^{(2)} + S_{int}^{(3)} + S_{int}^{(4)},\\
	S_{int}^{(1)} = \delta\int_{0}^{\beta}d\tau\int d\vec{r}\left\lbrace
   \psi_1 \sqrt{2}\rhozero (-\mu+g\rhozero)
      \right\rbrace,   \\
	S_{int}^{(2)} =\dsfrac{\delta}{2}\int_{0}^{\beta}d\tau\int d\vec{r} \left\lbrace \beta_1 \psi_1^2 +  \beta_2 \psi_2^2\right\rbrace ,  \\
	S_{int}^{(3)} = \dsfrac{\delta g\sqrt{2}\rhozero  }{2} \int_{0}^{\beta}d\tau\int d\vec{r} \left\lbrace
(\psi_1^2 +\psi_2^2)\psi_1
\right\rbrace, \\
	S_{int}^{(4)} =\dsfrac{\delta g}{8}\int_{0}^{\beta}d\tau\int d\vec{r}\left\lbrace \psi_1^4 +2\psi_1^2 \psi_2^2 +\psi_2^4 \right\rbrace,
\label{eqRSint(4)}
	\ea
\ee
where
\be
	\beta_1  = -\mu -X_1  +3g\rho_0, \quad \beta_2  = -\mu -X_2  +g\rho_0,
\ee
and $X_1$   and $X_2$  are the variational parameters, related to the normal $\Sigma_{n}$ and anomalous
$\Sigma_{an}$
self energies as $X_1=\Sigma_{n}+\Sigma_{an} -\mu$ and  $X_2=\Sigma_{n}-\Sigma_{an} -\mu$ . The free energy $\Omega$ can be evaluated as
\bea
\Omega = -T \ln Z(j_1, j_2)|_{j_1=0, j_2=0},
\eea
where the grand partition function is
\bea
Z(j_1, j_2)= e^{-S_0}\int D \psi_1 D \psi_2 e^{-\frac{1}{2}
	\int dx \int dx' \psi_a(x)G_{ab}^{-1}(x,x')\psi_b(x')} e^{-S_{int}}
e^ { \int dx [j_1(x)\psi_1(x)+j_2(x)\psi_2(x)] }
\label{eq:z}
\eea
in which we introduced  $x=(\tau,\vec{r})$ and $\int dx\equiv \int_{0}^{\beta}  d\tau\int d\vec{r}$.
For a uniform system, Green function is translationally invariant
\bea
G_{ab}(\vec{r}, \tau;  \vec{r}', \tau') = \frac{1}{V\beta}\sum_{n,k} e ^{i\omega_n(\tau-\tau')} e^{i\vec{k}(\vec{r}-\vec{r}')} G_{ab}(\vec{k}, \omega_n)
\label{eq: Gab}
\eea
with
\be
	\ba
	G_{11}(\vec{k}, \omega_n) =\dsfrac{\epsilon_k+X_2}{\omega_n^2+E_k^2},\\
	G_{22}(\vec{k}, \omega_n) =\dsfrac{\epsilon_k+X_1}{\omega_n^2+E_k^2},\\
	G_{12}(\vec{k}, \omega_n) =\dsfrac{\omega_{n}}{\omega_n^2+E_k^2},\\
	G_{21}(\vec{k}, \omega_n)  =-G_{12}(\vec{k}, \omega_n),\\
	E_k^2 =(\epsilon_k + X_1) (\epsilon_k + X_2)
\lab{greenf}	
\ea
\ee
where $\omega_n=2\pi n T$ is the Matsubara frequency.
 In the path integral formalism the expectation value of an operator $\langle\hat{O}(\ti{\psi}^{+},\ti{\psi})\rangle$ is defined as
\bea
\langle\hat{O}\rangle = \frac{1}{Z_0}\int D \ti{\psi}^{+} D \ti{\psi} \hat{O}(\ti{\psi}^+,\ti{\psi}) e^{-S(\ti{\psi}^{+},\ti{\psi})},
\eea
where $Z_0=Z(j_1=0, j_2=0, S_{int}=0)$ is the noninteracting partition  function.

 Particularly, using the well-known formula \re{formula} and following identities:
\be
 \ba
\ln Det[ G^{-1}]=\summa_{n,k}\ln (E_{k}^{2}+\omega_{n}^{2})=\summa_{k}[\beta E_k+2\ln (1-e^{-\beta E_k})],
\\
 \summa_{n=-\infty}^{n=\infty} \frac{1}{(\omega_{n}^{2}+E_{k}^{2})}=
\frac{\beta}{2E_k} \coth (\beta E_k/2),
 \lab{Aident}
\ea
\ee
one may show that \cite{ourkleinert}
{
\be
	\ba
	\langle \hat{O}(\psi_a(x)\psi_b(y))\rangle =\hat{O} \left( \dsfrac{\delta}{\delta j_a(x)}, \dsfrac{\delta}{\delta j_b(y)}\right)  exp{\left[ \frac{1}{2} \int j_a(x) G_{ab}(x,y)j_b(y)dx dy\right] }, \\
\langle \psi_{a}(x)\rangle=0, \quad \langle \psi_{a}(x)\psi_{b}(x')\rangle = G_{ab}(x,x'),\\
	\langle \psi_1(x)\psi_2(x) \rangle   = G_{12}(0) = \frac{1}{\beta}\dssum_{n}G_{12}(\vec{k}, \omega_n) = \frac{1}{\beta} \dssum_{n=-\infty}^{\infty}\frac{\omega_{n}}{\omega_{n}^2 + E_k^2}=0,\\
	\langle\psi_a^4(x)\rangle  = 3 G_{aa}^2(0),\\
	\langle\psi_1^2 (x) \psi_2^2 (x)\rangle  = G_{11}(0) G_{22}(0),\\
	G_{ab}(0) \equiv\dsfrac{1}{V\beta} \sum_{k,n} G_{ab}(k, \omega_{n}),\\
	\langle \psi_{a_1}, \psi_{a_2}\ldots\psi_{a_n}\rangle =0, \quad n=1,3,5 \ldots
\lab{a10}
	\ea
\ee
}
We now expand $exp(-S_{int})$ in Eqs.\,(A3) in powers of $\delta$
\be
e^{-S_{int}}= 1-S_{int}^{(1)} - S_{int}^{(2)} - S_{int}^{(3)} - S_{int}^{(4)}  +O(\delta^{2})
\lab{a11}
\ee
Expressing the ``noninteracting" partition function as
\be
\ba
Z_0(j) =\int D \psi_1 D \psi_2 e^{-\frac{1}{2}\int dx dx'
	\psi_a(x)G_{ab}^{-1}(x,x')\psi_b(x')} e^{\int dx j_a(x)\psi_a(x)}\\
 =(\sqrt{DetG}) \exp
{\left[\frac{1}{2}\int dx dx' j_a(x) \bar{G}_{ab}(x,x')j_b(x') \right]}
\lab{z0}
\ea
\ee
where $\bar{G}_{ab}(x,y)=[G_{ab}(x,y)+G_{ba}(y,x)]/2$,  one may obtain
\bea
Z(j) = e^{-S_0} \left[Z_0(j)-\langle S_{int}^{(1)}\rangle - \langle S_{int}^{(2)}\rangle - \langle S_{int}^{(3)} \rangle- \langle S_{int}^{(4)}\rangle
 \right]
\label{eq:z(j)}
\eea
where $\langle\hat{O}\rangle=[\int  D \psi_1 D \psi_2 e^{-S_{free}}\hat{O}(\psi_1, \psi_2) ]/Z_0(j)|_{(j=0)} $ and $Z_0(j)|_{(j=0)}= 1/\sqrt{{\rm Det}\,G^{-1}}$. The expectation values in (\ref{eq:z(j)}) can be easily calculated by
using Eqs.\, \re{a10}  as
\be
	\ba
	\langle S_{int}^{(1)}\rangle  =0, \quad  \langle S_{int}^{(3)}\rangle =0, \\	
	\langle S_{int}^{(2)}\rangle  = \dsfrac{1}{2}\int dx{(\beta_1 G_{11}(0) + \beta_2 G_{22}(0))}
	 =\dsfrac{\beta}{2}({\beta_1 B +\beta_2 A})\\
	\langle S_{int}^{(4)}\rangle  = \dsfrac{g\beta}{8}\left[ 3G_{11}^2(0)+3G_{22}^2(0)+2G_{11}(0)G_{22}(0)\right]
	 = \dsfrac{g\beta}{8}[3B^2 + 3A^2+2A B].
\lab{a15}
	\ea
	\ee
where $A=V(\rho_1-\sigma)$, $B=V(\rho_1+\sigma)$.
Thus, using the formula $\ln(1+x)\approx x$ we obtain
\bea
\Omega = -T \ln Z(j)|_{j=0} = -T \ln e^{-S_0} -T \ln Z_0 + T\langle S_{int}^{(2)}\rangle + T\langle S_{int}^{(4)}\rangle
\label{eq: finalomega}
\eea
where we set $\delta=1$. Finally, using \re{a15} gives
\be
	\ba
	\Omega = \Omega_0 + \Omega_{free} + \Omega_2 + \Omega_4 \\
	\Omega_0  = -\mu V\rho_0 + \dsfrac{gV\rhozero^2}{2},\\
	\Omega_{free}  = \frac{1}{2} \sum_{k} (E_k -\epsilon_k) + T \sum_{k}ln(1-e^{-\beta E_k}),\\
	\Omega_2  =  \frac{1}{2} [\beta_1 B + \beta_2 A]\\
	\Omega_4  =\dsfrac{g}{8V}[3A^2 +3B^2 +2AB],
	\ea
\ee
In above equations  $X_1\equiv2\Delta$ can be found from equation $\partial \Omega/\partial X_1=0$
which leads to MFA equation \re{delta}. As to $X_2$ it should be  set to zero, $X_2=0$, in order to
make the dispersion  similar to  the Bogoliubov one: $E_k=\sqrt{\veps_k}\sqrt{\veps_k+2\Delta}$ in accordance with
 Hugenholtz - Pines theorem \ci{pinestheorema}. As a result, one obtains
 \be
 \ba
 X_1=2g(\rhozero+\sigma)=2g\rho+2g(\sigma-\rhoone),\\
 \mu=g\rho+g\rhoone-g\sigma.
 \ea
 \ee

\clearpage
\section{Summary of main equations in various approaches of MFT}
\label{sec:B}
We present  the total energy of a Bose system at $T=0$ as:
\be
E=\frac{V\rho^2 g}{2}(1+ \tilde{E}_0) +\frac{8Vm^{3/2}}{15\pi^2}\tilde{E}_{fluc}
\lab{appe}
\ee
where, $\tilde{E}_0$ and  $\tilde{E}_{fluc}$ are shown on Table 1 (columns IV, V).
Tan's contact, calculated from any of equations \re{cn}-\re{cpsi} may be simply presented as
$C_{x}=16\pi^2a^2\rho^2(1+W_x)$, ($x=n, E, \psi$)  where $W_x$ are given on columns VI-VIII of Table 1.
The second column of this table includes equations for the condensed fraction $\rhozero$
and for the reduced self energy $\Delta=(\Sigma_{n}+\Sigma_{an}-\mu)/2$. Note that, $ \rhozero  $ is fixed
as $\rhozero=n_0\rho=\rho(1-8\sqrt{\gamma}/3\sqrt{\pi})$ in Bogoliubov approximation, while it should be numerically evaluated as
solutions of MFT equations in other approaches.

\begin{sidewaystable}
\caption{  MFT equations, the total energy and Tan's contact in MFT}
\renewcommand{\arraystretch}{1.3}
\begin{tabular}{|l| l| l|l|l|l|l|l|}
\hline
 MFT app. & $\rhozero$ and MFT equations &Dispersion       &$\tilde{E}_0$ & $\tilde{E}_{fluc}$  &$W_n$ &$W_E$ &$W_\psi$  \\
 \hline
HFB
&
 $
\ba
\rhozero=\rho-\rhoone\\
\Delta=g(\rhozero+\sigma) \\
\rhoone=(\Delta m)^{3/2}/3\pi^2\\
  \sigma=m^{3/2}\Delta \sqrt{g\rhozero}/\pi^2\\
\ea
$
&
$E_k=\sqrt{\veps_k}\sqrt{\veps_k+2\Delta}$
&
$\tilde{E}_0=n_{1}^{2}-\sigtil^2-2n_1\sigtil$
&
$\Delta^{5/2}$
 &
  $
  \ba
  (n_1-\sigtil)*\\
  (n_1-\sigtil-2)
  \ea
  $
&
 $
\ba
n_\sigma+\Delta'_{a}\left[\frac{2mn_1}{\pi\rho}+\dsfrac{3an_\sigma}{\Delta}\right]\\
n_\sigma=n_{1}^2-\tilde\sigma^2-2n_1\tilde\sigma\\
\Delta'_{a}=\frac{\Delta}{a}\frac{1}{[1+6\pi a (\rhoone-\sigma)/m\Delta]}
\ea
$
&
$
\ba
2(n_1+\tilde\sigma-\\2n_1 \tilde\sigma)-
n_\sigma
\ea
$
   \\
\hline
Gaussian
&
 $
\ba
{\tilde{n_0}}^3+p{\tilde{n_0}}^2-p=0,\\
{\tilde{n_0}}=\sqrt{n_0}, \\
 p=3\sqrt{\pi}/8\sqrt{\gamma}
\ea
$
 &
 $E_k=\sqrt{\veps_k}\sqrt{\veps_k+2g\rhozero}$
 &
  0
 &
 $(g\rhozero)^{5/2}$ &$n_1(n_1-2)$
 &
  $\dsfrac{ 64 n_{0}^{5/2}\sqrt{\gamma}   }
 {3(\sqrt{\pi}+4\sqrt{\gamma n_0})}$
 &
 $\dsfrac{64n_0^{3/2}\sqrt{\gamma}}{3\sqrt{\pi}}$
\\
    \hline
Bogoliubov
&
 $
\ba
n_0=1-8\sqrt{\gamma}/3\sqrt{\pi}
\ea
$
 &
 $E_k=\sqrt{\veps_k}\sqrt{\veps_k+2g\rho}$
 &
 0
  &
  $(g\rho)^{5/2}$&1 &$\dsfrac{64\sqrt{\gamma}}{3\sqrt{\pi}}$
&
$\dsfrac{64\sqrt{\gamma}}{3\sqrt{\pi}}$
  \\
 \hline
\end{tabular}

\end{sidewaystable}

\newpage

\bb{99}
\bi{anderson1995} Anderson, M. H., J. R. Ensher, M. R. Matthews, C. Wieman,
and E. A. Cornell,  Science 269, 198  (1995)
\bi{Andersen} J. O. Andersen,
 Rev. Mod. Phys. \textbf{76}  599  (2004)
\bi{tutorial}  N.P. Proukakis, B. Jackson, J. Phys. B: At. Mol. Opt. Phys. {\bf 41}  203002 (2008)
\bi{Yukalovann}  V.I. Yukalov, Ann. Phys. {\bf 323}  461  (2008)
\bi{tan1} S. Tan, Ann. Phys. (N.Y.) 323, 2952 (2008).
\bi{tan2} S. Tan, Ann. Phys. (N.Y.) 323, 2971 (2008).
\bi{tan3} S. Tan, Ann. Phys. (N.Y.) 323, 2987 (2008).
\bi{brprl100} E. Braaten and L. Platter Phys. Rev. lett. { \bf 100}, 205301 (2008)
\bi{brprl104} E. Braaten, D. Kang, and L. Platter, Phys. Rev. Lett. {\bf 104}
223004 (2010).
\bi{brprl106} E. Braaten, Daekyoung Kang,  and L. Platter, Phys. Rev. lett. {\bf 106}, 153005 (2011)
\bi{lang2017} G. Lang,  P. Vignolo, and A. Minguzzi, Eur. Phys. J. Special Topics { \bf 226}, 1583 (2017)
\bi{werner}  F. Werner, and Y. Castin, Phys.  Rev. {\bf A 86}, 053633 (2012)
\bi{combescot} R. Combescot, F. Alzetto and X. Leyronas, Phys.  Rev. {\bf A 79}, 053640 (2009)
\bi{pitbook} L. Pitaevskii and S. Stringari , {\it Bose- Einstein Condensation and Superfluidity  }
, Oxford University Press, 2015
\bi{ourjt}  A. Rakhimov, M. Nishonov, and  B. Tanatar, Phys. Lett. {\bf A384}  126313 (2020)
\bi{sagi} Y. Sagi, T. E. Drake, R. Paudel, and D. S. Jin, Phys. Rev. Lett. {\bf  109}, 220402 (2012)
\bi{stewart} J. T. Stewart, J. P. Gaebler T. E. Drake and D. S. Jin , Phys. Lett. {\bf 104} 235301 (2010)
\bi{wild} R. J. Wild, P. Makotyn, J. M. Pino, E. A. Cornell, and D. S. Jin
Phys. Rev. Lett. {\bf 108} , 145305 (2012)
\bi{makotyn} P. Makotyn, C.  E. Klauss, D. L. Goldberger, E. A. Cornell, D. S. Jin
 Nature Physics {\bf 10}, 116 (2014)
\bi{chang} R. Chang, Q. Bouton, H. Cayla, C. Qu, A. Aspect,  C. I. Westbrook, and D. Clement,
Phys. Rev. Lett. {\bf 117}, 235303 (2016)
\bi{fletcher}  R. J. Fletcher, R. Lopes, J. Man, N. Navon, R. P. Smith,
 M. W. Zwierlein, Z. Hadzibabic, Science {\bf  355} ,  377   (2017)
\bi{faddeev}  L. D. Faddeev, A. A. Slavnov
{\it Gauge fields, introduction to quantum theory}. Perseus Books, 1991.
\bi{ouryee}   A. Rakhimov, C. K. Kim, S.-H. Kim, and J. H. Yee,
 Phys. Rev. A \textbf{77}  033626   (2008).
\bi{stevenson} P. M. Stevenson  Phys. Rev. D 32, 1389, (1985).
\bi{pinto} F. F. de Souza Cruz, M. B. Pinto, and R. O. Ramos,
 Phys. Rev. B \textbf{64},  014515  (2001)
\bi{stancu} I. Stancu and P. M. Stevenson,
Phys. Rev. D \textbf{42}  2710,  (1990).
\bi{bog}  N.N. Bogoliubov, J. Phys. USSR, {\bf 11}, 23 (1947)
\bi{leeyang}  T. D. Lee, Kerson Huang, and C. N. Yang, Phys. Rev.
106, 1135 (1957).
\bi{schakel}  A. M. J. Schakel   arXiv:1007.3452
\bi{petrov}  D. S. Petrov. Phys. Rev. Lett. {\bf 115}, 155302 (2015).
\bi{rossisalas}  M. Rossi and L. Salasnich, Phys. Rev. {\bf A 88} , 053617 (2013)
\bi{brnieto} E. Braaten and A. Nieto, Eur. Phys. J. B11, 143 (1999).
\bi{ouraniz2part1}  A. Rakhimov, A. Khudoyberdiev, L Rani, and B. Tanatar, arXiv:1909.00281
\bi{ourkleinert}  H. Kleinert, Z. Narzikulov, and A. Rakhimov,
 J. Stat. Mech. P\textbf{01003}  1742, (2014).
\bi{pinestheorema}    N. M. Hugenholtz and D. Pines,
  Phys. Rev. \textbf{116 } 489506    (1959).
 \eb

\edc